# Near constant loss dielectric response in $2Bi_2O_3$-$B_2O_3$ glasses

G Paramesh and K B R Varma[*]

Materials Research Center

Indian Institute of Science

Bangolore-560012

India.

E-Mail: kbrvarma@mrc.iisc.ernet.in

**Abstract**. Electrical conduction and relaxation phenomena in bismuth borate glasses in the composition $2Bi_2O_3$-$B_2O_3$ ($Bi_4B_2O_9$) were investigated. Dielectric studies carried out on these glasses revealed near constant loss (NCL) response in the 1 kHz to 1MHz frequency range at moderately high temperatures (300-450K) associated with relatively low loss (D=0.006) and high dielectric constant ($\varepsilon_r'$=37) at 1kHz,300K. The variation in AC conductivity with temperature at different frequencies showed a cross over from NCL response characterized by local ion vibration within the potential well to universal Joncher's power law dependence triggered by ion hopping between potential wells or cages. Thermal activation energy for single potential well was found to be 0.48±0.05eV from cross over points. Ionic conduction and relaxation processes were rationalized by modulus formalism. The promising dielectric properties (relatively high $\varepsilon_r'$ and low D) of the present glasses were attributed to high density (93 % of its crystalline counterpart), high polarizability and low mobility associated with heavy metal cations, $Bi^{3+}$.

## 1. Introduction:

Heavy metal oxide containing glasses have been of technological interest for a possible use in the field of nonlinear optics and infrared optics owing to their higher density, lower field strength and higher polarizabilty [1, 2]. These exhibit high refractive indices accompanied by high hyperpolarizabity which result in high third order nonlinear susceptibilities. Among heavy metal oxide glasses bismuth based glasses especially bismuth borates were reported to be important in the area of linear and nonlinear optics [3-5]. In the binary bismuth borate system, $nBi_2O_3-mB_2O_3$ the formations of many stable single crystalline phases along with their physical properties were reported [6]. Among them $Bi_2B_8O_{15}$ and $BiB_3O_6$ are promising for nonlinear optical applications [7]. The new, stable bismuth borate single crystal $Bi_4B_2O_9$ which has high concentration of $Bi_2O_3$ was reported to possess attracting optical properties among the hitherto reported bismuth borate crystals [8]. Investigations into functional materials in their glassy state and glasses comprising nano/micro crystals are of both scientific and technological importance as these can be alternatives to their single crystalline counterparts. $Bi_4B_2O_9$ (in $2Bi_2O_3-B_2O_3$ system) is considered to be important for applications in the infrared wavelength range and for third order harmonic generation. Since the dielectric and associated electrical properties (though in optical frequency regime) have direct influence on the electro-optic and non-linear optical characteristics of these materials, we felt that it is important to characterize them for their basic electrical properties to begin with. The present article describes the dielectric properties of these glasses over 1 kHz-1 MHz frequency range as a function of temperature (300-623K). Interestingly the dielectric constant has been found to be relatively higher as ever reported for glasses of the above kind and showed frequency independent dielectric response. The details of which are reported in the following sections.

## 2. Experimental:

Transparent glasses in the system $2Bi_2O_3$-$B_2O_3$ were fabricated via the conventional melt quenching technique. For this $Bi_2O_3$ and $H_3BO_3$ were weighed (for 10 gm.) in appropriate molar ratio and mixed thoroughly. The mixture was melted in a platinum crucible using Lenton furnace at 1073K/30 min. Then the melt was quenched to room temperature by pouring it on a steel plate and subsequently pressing it with another one to obtain glasses of thickness 0.8-1mm. As quenched glasses were annealed for 5 h at 553K which is below the glass transition temperature to relieve the glasses from thermal stresses that are likely to be associated with them. Archimedes' principle was used to measure the density of the glasses using Xylene as a suspension liquid. The X-ray powder diffraction (Bruker D8) studies were carried out to confirm the amorphous nature of the as quenched samples. Differential scanning calorimetry (DSC, Perkin Elmer model) was used to obtain the glass transition and crystallization temperatures.

The capacitance and dielectric loss(D) measurements on the as-quenched (annealed) polished glass plates were performed using impedance gain phase analyzer (HP 4194 A) in the 1 kHz–1 MHz frequency range with a signal strength of 0.5 $V_{rms}$ at various temperatures (300–623K). Thin silver leads were bonded to the sample (that was gold sputtered) using silver epoxy for making electrical measurements. Based on the capacitance data, the dielectric constants were evaluated by taking the dimensions and electrode geometry of the sample into account.

**3. Results and Discussion:**

The amorphous nature of the as quenched yellowish transparent samples was confirmed by X-ray powder diffraction studies (figure1). The samples that were subjected to the differential scanning calorimetry (DSC) at a heating rate 15 K/min exhibited the endotherm at 611K followed by the intense exotherm at 694K corresponding to the glass transition and crystallization temperatures respectively and the corresponding thermo gram is shown in figure 2. The measured density of the glasses was 7.86 g/cc which is about 93% of the theoretical density of its single crystalline counterpart.

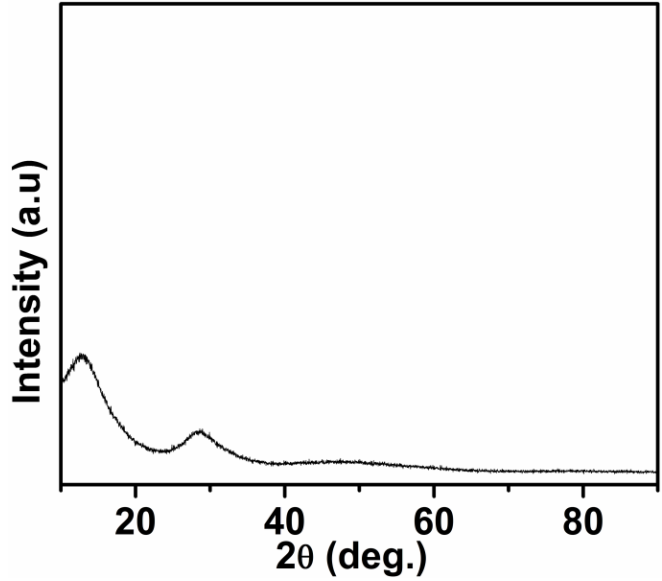

*Figure 1 X-ray diffraction pattern for the as-quenched $2Bi_2O_3$-$B_2O_3$ glasses.*

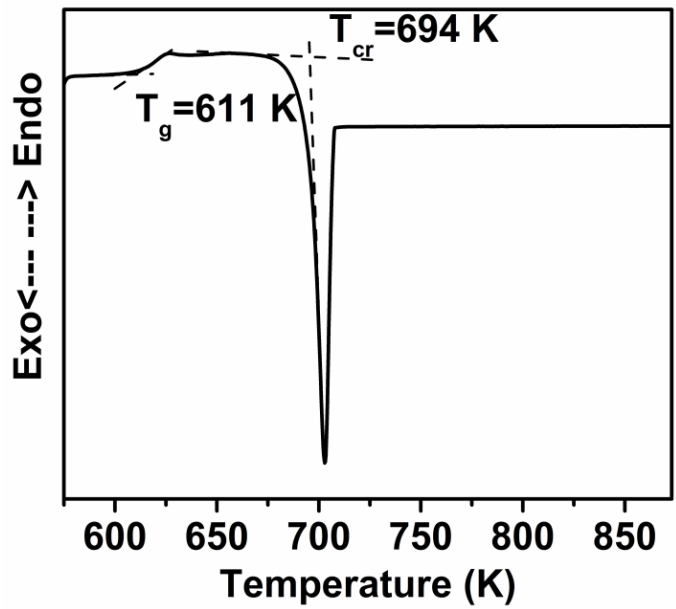

*Figure 2 DSC trace for the as-quenched glasses.*

3.1. *Dielectric Studies:*

The dielectric constant ($\varepsilon_r'$) and dielectric loss (D) as a function of frequency in the 1 kHz-1 MHz range at different temperatures are shown in figures 3(a) and 3(b). $\varepsilon_r'$ and D obtained respectively for the BBO glasses are 36±0.5 and 0.006±0.001 at 1 kHz, 300K and they were found to be almost independent of frequency.

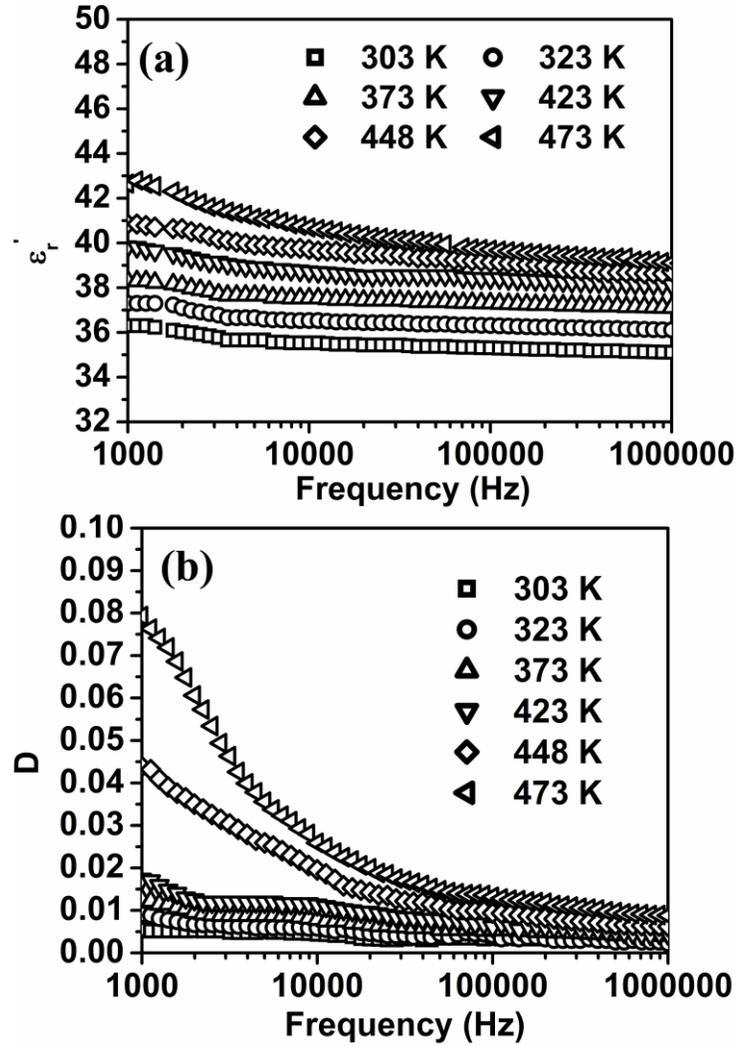

*Figure 3 Frequency dependent (a) dielectric constant and (b) dielectric loss at various temperatures.*

The low loss and relatively high dielectric constant values of the BBO glasses are attributed to the presence of higher content of $Bi_2O_3$ associated with high density. Heavy metal cations such as bismuth ($Bi^{3+}$) possess high polarizability owing to their low field strengths (large ionic radius) imparting high dielectric constant to the glasses.

The frequency independent nature of $\varepsilon_r'$ and D is referred to as the near constant loss (NCL) phenomenon. The NCL response had been observed in many dielectrics including glasses, ceramics at sufficiently low temperatures and high frequencies [9-12]. The incidence of NCL response at moderately higher temperatures (300K-450K) is the characteristic feature of the present BBO glasses and these investigations infer that NCL type response would also be active at moderately high temperatures depending on the physical system and its constituent ions. $\varepsilon_r'$ and D as a function of temperature 323-450K at various frequencies are also depicted in figure 4(a,b). Weak temperature dependence of ionic conduction type polarization is noticed.

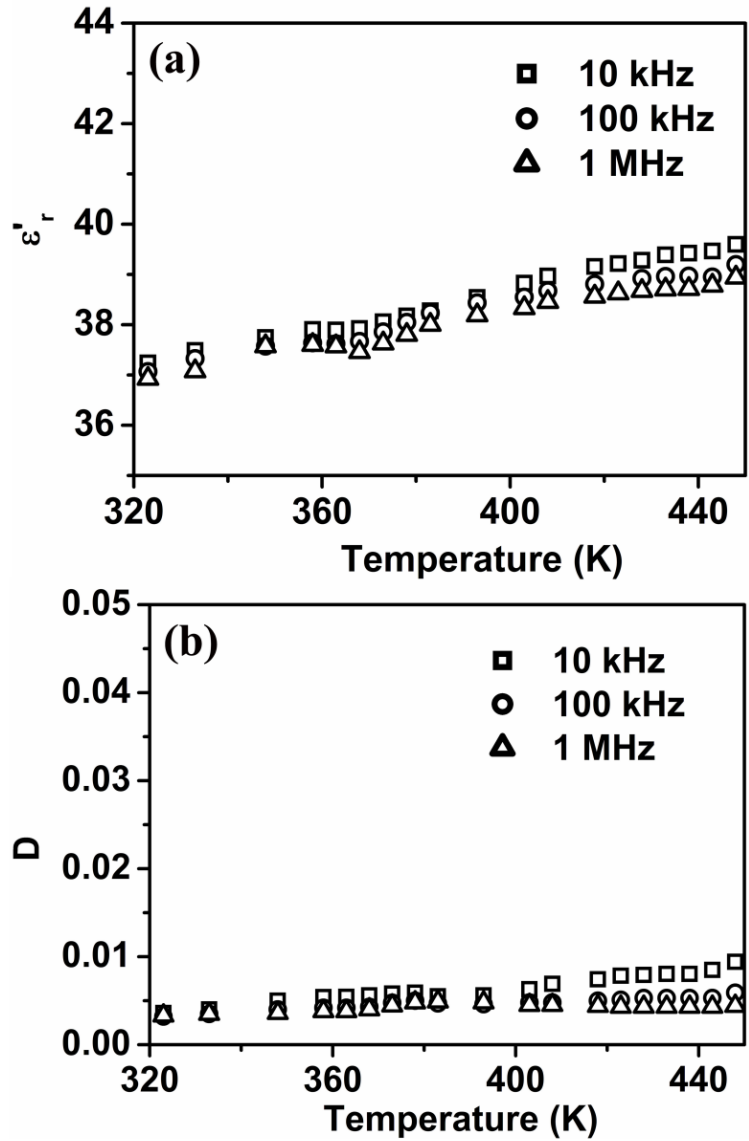

*Figure 4 Temperature dependent (a) dielectric constant and (b) loss at three different frequencies.*

## 3.2. Conductivity Studies:

AC conductivity ($\sigma_{AC}$) of glasses was calculated from dielectric data using the formula

$$\sigma_{AC} = \varepsilon_0 \varepsilon'_r D \omega \qquad (1)$$

Where $\varepsilon_0$ is the permittivity of the free space and $\omega$ is the angular frequency. Variation of $\sigma_{AC}$ (log scale) with temperature (linear scale) at different fixed frequencies is shown in figure 5.

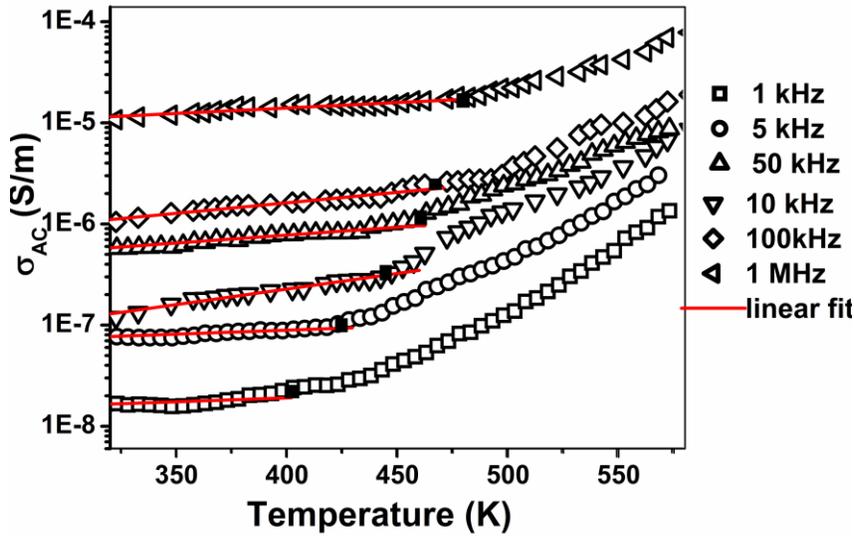

*Figure 5 AC conductivity vs. temperature plots at various frequencies. Solid squares are the symbols distinguishing the NCL response and power law regime. Solid lines are theoretical fits to the exponential temperature dependent conductivity in NCL regime*

At higher temperatures the conductivity curves at different frequencies are found to tend to merge into a single curve due to the dominant DC conduction which is due to a long range motion of ions that are activated thermally. The conductivity decreases with the decrease in temperature and follows fractional exponent power law (Joncher's law) of frequency due to hopping of ions from one potential well to the other which is a short range motion in nature. But at sufficiently low temperatures it has been observed in many systems that the conductivity fractional power law of frequency is no longer valid and rather it varies linearly with frequency (NCL response). This is also noticed at high frequencies at moderate temperatures. In the case of present BBO glasses the NCL response is found to emerge at moderate temperatures. The cross over temperature points between Joncher's response and NCL response are symbolized as solid squares in figure 5 (below 450K). This regime is characterized by exponential temperature dependence of conductivity (exp (bT) where b is constant and T is temperature) [13, 14] and the solid lines in figure 5 are the theoretical fits to exp(bT). The fitting parameter 'b' obtained for the present system is $0.0019\pm0.0005$.

This implies that ionic species involved are confined or caged within their potential wells with local vibrational motion below cross over temperature points. The ion should cross its own local potential barrier or cage so that it can hop to other site. The thermal activation energy required for ions to participate in hopping between cages is calculated from cross over points using Arrhenius plot which is shown in figure 6.

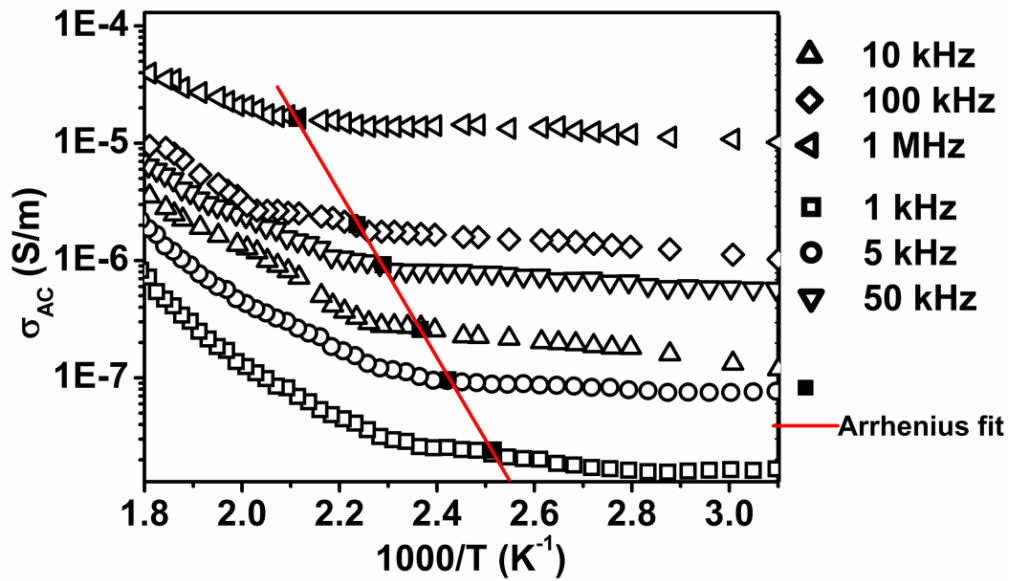

*Figure 6 Arrhenius plots of AC conductivity at different frequencies. Solid squares are cross over points and solid line is the linear fit.*

It is found to be $0.48\pm0.05$ eV which is less than the activation energy required for long range motion of DC conduction calculated from modulus formalism the details of which are illustrated in the subsequent section.

The NCL behaviour was described in the literature in the form of augmented Joncher's law since it emanates at sufficiently high frequencies at all the temperatures and the resulting modified fractional power law is given as follows[15]

$$\sigma_{AC} = \sigma_{DC} + A\omega^s + B\omega \tag{2}$$

However because of the presence of cross over temperature in the present case in 1 kHz- 1 MHz frequency window, we followed the two term Joncher's law

$$\sigma_{AC} = \sigma_{DC} + A\omega^s \tag{3}$$

and fitted to the experimental data at different temperatures. Frequency dispersion plots of $\sigma_{AC}$ at various temperatures are shown in figure 7 for BBO glasses. Solid lines are theoretical fitted curves to (3). The power law exponent (s) obtained from the fitting is found to be very close to 1 in the 300-450K temperature range (i. e., NCL regime). Over this range the deviation of 's' from unity is noticeable and its variation with temperature is shown in the inset of figure 7. The deviation of the 's' from unity with increasing temperature also suggests that there is an increasing interaction among ionic species which is expected when the hopping phenomenon dominates.

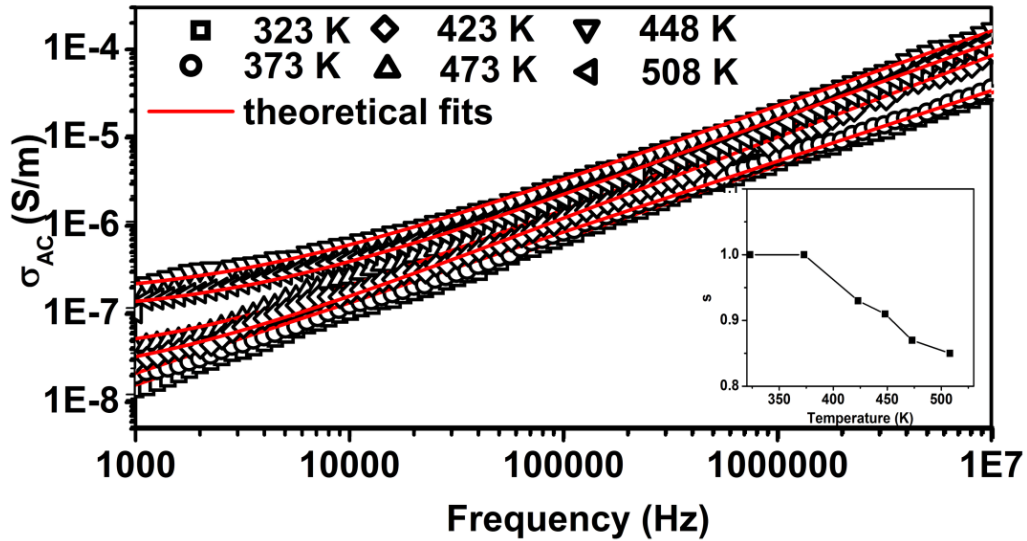

*Figure 7 AC conductivity dispersion curves at various temperatures (solid lines are theoretical fits to power law of conductivity) and inset showing the variation of power law exponent (s) with temperature.*

3.3. *Electric Modulus Studies:*

The electric modulus formalism was invoked to rationalize the relaxation process in the present glass system. Electric modulus formalism is advantageous to extract the bulk response of the material as it suppresses the electrode polarization and masks the relaxation response due to conduction. The electric modulus is defined as [16]

$$M^* = M' + iM'' = \frac{1}{\varepsilon^*} = \frac{\varepsilon'}{\varepsilon'^2 + \varepsilon''^2} + i\frac{\varepsilon''}{\varepsilon'^2 + \varepsilon''^2} \qquad (4)$$

where M', M" and $\varepsilon'$, $\varepsilon''$ are the real and imaginary parts of modulus and dielectric constants respectively. The M' and M" plots in the frequency range under study at various temperatures are shown in figure 8. One could see that the M' (figure 8 (a)) tends to zero in the low frequency regime at all the temperatures under study which is due to electrode polarization. At higher frequencies it reaches $M'_\infty (=1/\varepsilon_\infty)$ which is the characteristic relaxation. The M" plots in the figure 8(b) at different temperatures exhibited clear relaxation peaks at characteristic relaxation frequencies and the peak is found to shift to higher frequencies with increasing temperature suggesting the association of thermally activated phenomenon in relaxation. The frequency regime that is below the M" peak position indicates the range with which the ions drift to long distances. In the frequency range which is above that of the peak, the ions are spatially confined to potential wells and free to move within the wells. The frequency range where the peak occurs is suggestive of the transition from long-range to short-range mobility. The absence of the relaxation peaks in NCL regime may be noted. The characteristic feature of the NCL response in the present glasses is mainly due to the absence of mobile cations such as Li and Na (relax at relatively higher frequencies). The active mobile cation in the glass matrix is Bi. Bi cations migrate through broken network of non-bridging oxygens with ease. Bi ion conduction reported to be active in the bismuth silicate and bismuth germinate glasses [17]. The low mobility, heavy metal $Bi^{3+}$ cations would relax at relatively lower frequencies while interacting with oscillating electric field at various frequencies and possess higher thermal activation energy. As the temperature increases the relaxation frequency shifts towards higher frequency side due to thermal activation and one could see the characteristic peak at a certain temperature.

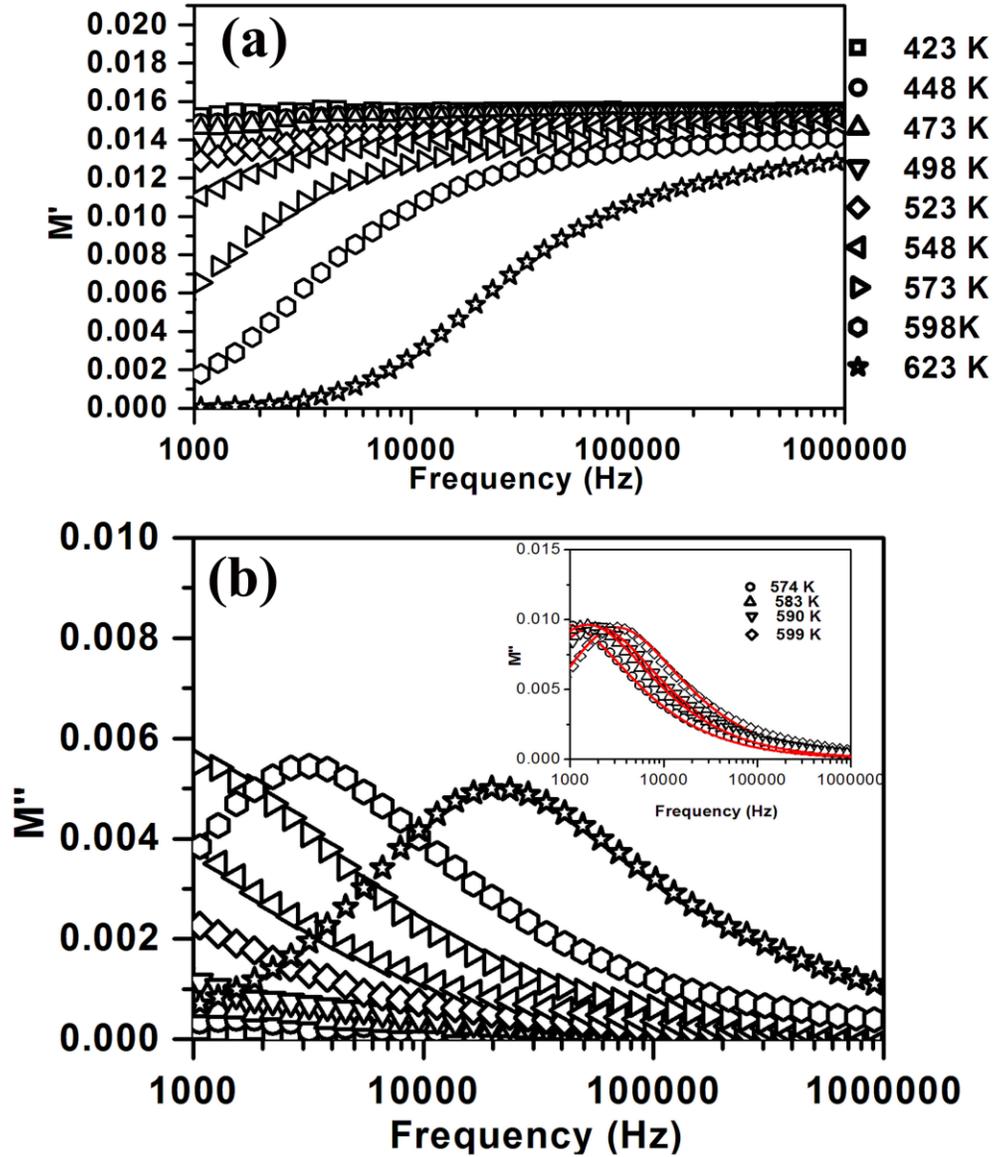

*Figure 8 Frequency dependent (a) real and (b) imaginary parts of electric modulus at various temperatures (inset showing theoretical fitted (solid lines) curves to the imaginary electric modulus curves).*

The electric modulus could be expressed as the Laplace transform of a relaxation function $\phi(t)$ [18]

$$M^* = M_\infty \left[ 1 - \int_0^\infty \exp(-\omega t)\left(-\frac{d\phi}{dt}\right) dt \right] \tag{5}$$

where the function ϕ(t) is the time evolution of the electric field within the materials and is usually taken as the Kohlrausch-Williams-Watts (KWW) function [19]

$$\phi(t) = \exp\left[-\left(t/\tau_m\right)^\beta\right] \qquad (6)$$

where $\tau_m$ is the conductivity relaxation time and the exponent $\beta$ (0 1] indicates the deviation from Debye type relaxation. The value of $\beta$ could be determined by fitting the experimental data in the above equations. But it is desirable to reduce the number of adjustable parameters while fitting the experimental data. Keeping this point in view, the electric modulus behavior of the present glass system is rationalized by invoking modified KWW function suggested by Bergman. The imaginary part of the electric modulus (M") is defined as [20]

$$M'' = \frac{M''_{max}}{(1-\beta) + \frac{\beta}{1+\beta}\left[\beta(\omega_{max}/\omega) + (\omega/\omega_{max})^\beta\right]} \qquad (7)$$

where $M''_{max}$ is the peak value of the M" and $\omega_{max}$ is the corresponding frequency. The above equation (7) is effectively described for $\beta \geq 0.4$. Theoretical fits of Eq. 7 to the experimental data are illustrated in the inset of figure 8 (b) as the solid lines. The experimental data are well fitted to this model except in the high frequency regime. From the fitting of M" versus frequency plots, the value of $\beta$ is determined. The value of $\beta$ is found to be 0.71±0.03 in the 573K–623K temperature range indicating narrow distribution of relaxation times owing to low concentration of conducting species in the present glasses. It also reflects the local structural environment in homogeneous glasses [21].

It is of interest to investigate into the transport mechanism in the present glasses. Therefore, the dc conductivity at different temperatures $\sigma_{DC}(T)$ was calculated from the electric modulus data [14]. The dc conductivity could be extracted using the expression

$$\sigma_{DC}(T) = \frac{\varepsilon_o}{M_\infty(T) * \tau_m(T)} \left[ \frac{\beta}{\Gamma(1/\beta)} \right] \quad (8)$$

where $\varepsilon_o$ is the free space dielectric constant, $M_\infty(T)$ is the reciprocal of high frequency dielectric constant, and $\tau_m(T)$ is the temperature dependent relaxation time. Figure 9 shows the dc conductivity data obtained from the above expression (8) at various temperatures.

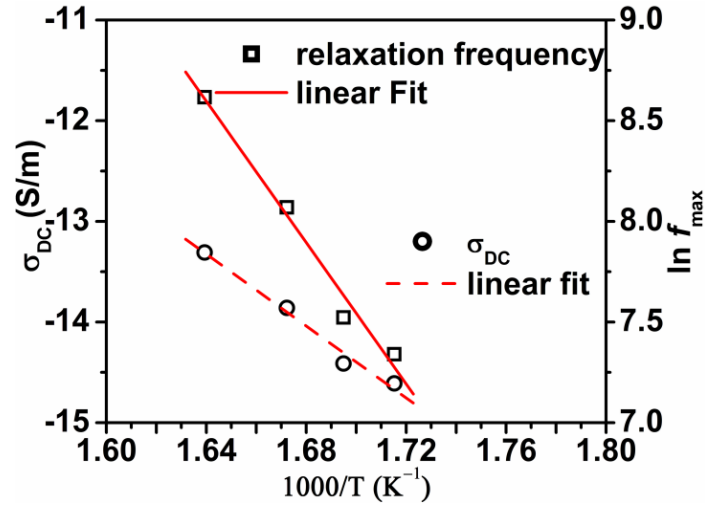

*Figure 9 Plots of $\ln\sigma_{DC}$ and $\ln f_{max}$ vs. 1000/T (solid lines are Arrhenius linear fits).*

The activation energy for the dc conductivity is calculated from the plot of ln $\sigma_{DC}$ versus $1000/T$ (figure 9) for these glasses. The plot is found to be linear and fitted using the following Arrhenius equation

$$\sigma_{DC}(T) = B \exp\left(-\frac{E_{DC}}{kT}\right) \quad (9)$$

where $B$ is the pre-exponential factor, $E_{DC}$ is the activation energy for the dc conduction. The activation energy is calculated from the slope of the fitted line and found to be 1.48±0.05 eV. The relaxation frequency associated with the above process is also determined from the plot of M" versus frequency. The activation energy involved in the relaxation process of ions could be obtained from the temperature dependent relaxation frequency

$$f_{max} = f_o \exp\left(\frac{-E_R}{kT}\right) \tag{10}$$

where $E_R$ is the activation energy associated with the relaxation process, $f_o$ is the pre-exponential factor, $k$ is the Boltzmann constant, and $T$ is the absolute temperature. Figure 9 shows a plot between ln $f_{max}$ and 1000/$T$ along with the theoretical fit (solid line) to the above equation (10). The value that is obtained for $E_R$ is 1.5±0.06 eV, which is in close agreement with that of the activation energy associated with dc conductivity. It suggests that similar energy barriers are involved in both the relaxation and conduction processes. This activation energy is large as compared to the single well potential barrier energy which is 0.48±0.05 eV. The activation energy for the relaxation or conduction is sum of this single well potential barrier and the barrier between two neighbouring potential wells.

## 4. Conclusions:

The glasses of composition $2Bi_2O_3$–$B_2O_3$ obtained via conventional melt quenching were found to have higher values of dielectric constant with relatively low loss. Frequency independent dielectric response at moderately higher temperatures and the presence of thermal crossover between the near constant loss (NCL) response and Jonchers's universal relaxation in the experimental frequency and temperature range under study have been some of the important features of the present glasses. These features were attributed to the response of heavier $Bi^{3+}$ cations to the oscillating electrical signal at different temperatures.